\documentclass[usenatbib]{basi}

\usepackage[british]{babel}
\usepackage[varg]{txfonts}
\usepackage{rotating}
\usepackage{dcolumn}

\begin{document}
\title[Black hole transients]{Black hole transients}

\author[T.~M.~Belloni, S.~E.~Motta and T.~Mu\~noz-Darias]%
       {T.~M.~Belloni,$^1$\thanks{email: \texttt{tomaso.belloni@brera.inaf.it}}
       S.~E.~Motta$^{1,2}$ and T.~Mu\~noz-Darias$^{3,1}$\\
       $^1$INAF-- Osservatorio Astronomico di Brera, Via E. Bianchi 46, I-23807 Merate, Italy\\
       $^2$Universit\`a dell'Insubria, Via Valleggio 11, I-22100 Como, Italy\\
       $^3$Instituto de Astrof\'isica de Canarias, Calle V\'ia L\'actea s/n, 38205 La Laguna, Tenerife, Spain}

\pubyear{2011}
\volume{39}
\pagerange{\pageref{firstpage}--\pageref{lastpage}}

\date{Received 2011 August 23; accepted 2011 September 12}

\maketitle
\label{firstpage}

\begin{abstract}
Sixteen years of observations of black hole transients with the Rossi X-ray Timing Explorer, complemented by 
other X-ray observatories and ground-based optical/infrared/radio telescopes have given us a clear view of 
the complex phenomenology associated with their bright outbursts. This has led to the definition of a small 
number of spectral/timing states which are separated by marked transitions in observables. The association 
of these states and their transitions to changes in the radio emission from relativistic radio jets completes
the picture and have led to the study of the connection between accretion and ejection. A good number of 
fundamental questions are still unanswered, but the existing picture provides a good framework on which to 
base theoretical studies. We discuss the current observational standpoint, with emphasis onto the spectral 
and timing evolution during outbursts, as well as the prospects for future missions such as ASTROSAT (2012) 
and LOFT ($>$2020 if selected). 
\end{abstract}



\section{Introduction}\label{s:intro}

The first black hole transient (BHT), A 0620$-$00, was discovered more than thirty years
ago \citep{1975Natur.257..656E} and already showed that the spectral evolution of these objects 
was very complex.  The first major step forward came with the Japanese satellite Ginga, which thanks
to the presence of an all-sky monitor and a large pointed detector led to the 
discovery and detailed study of a few objects \citep{1995xrbi.nasa..126T}.
Complex variability patterns were discovered, including Quasi-Periodic Oscillations,
which led to the extension of the hard/soft states of black hole binaries (originally
developed for Cygnus X-1) to additional states \citep[see e.g.,][]{1993ApJ...403L..39M}.
Because of the scarcity of sources and of coverage, Ginga
data gave only a limited view of the spectral/timing evolution of BHTs. 
Around the same epoch, optical observations yielded the first strong dynamical evidence for a 
BH (i.e. compact object heavier than $\sim$3~M$_{\odot}$) in A 0620$-$00 \citep{1986ApJ...308..110M}, 
which was fully confirmed a few years later by \cite{1992Natur.355..614C} with the measurement of a mass 
function $>$6 M$_{\odot}$ in another BHT, the Ginga source GS 2023+338 (=V404 Cyg).
The situation changed
with the launch of the Rossi X-ray Timing Explorer (RXTE) at the end of 1995. Sixteen years
of operations (with high level of flexibility and fast response) led to the observation of a large number 
of objects through extensive campaigns, 
often with coordinated observations at other wavelengths \citep[see e.g.,][]{2006ARA&A..44...49R}.
From this wealth of data, clear patterns emerged which are now leading to the development of
theoretical models. The evolution of spectral, timing and spectral/timing parameters can now be
classified into a small number of states which can be easily identified and followed throughout an
outburst \citep[see][]{2010LNP...794...53B}, as well as linked to observations at lower energies 
\citep[see][]{2010LNP...794..115F}.
In this paper, we summarize the current observational status, focussing on the outburst evolution 
and spectral states, and briefly discuss the contributions from upcoming missions like ASTROSAT and LOFT.

\section{The tools of the trade: fundamental diagrams}

\begin{sidewaysfigure}
     \centerline{\includegraphics[width=19.0cm]{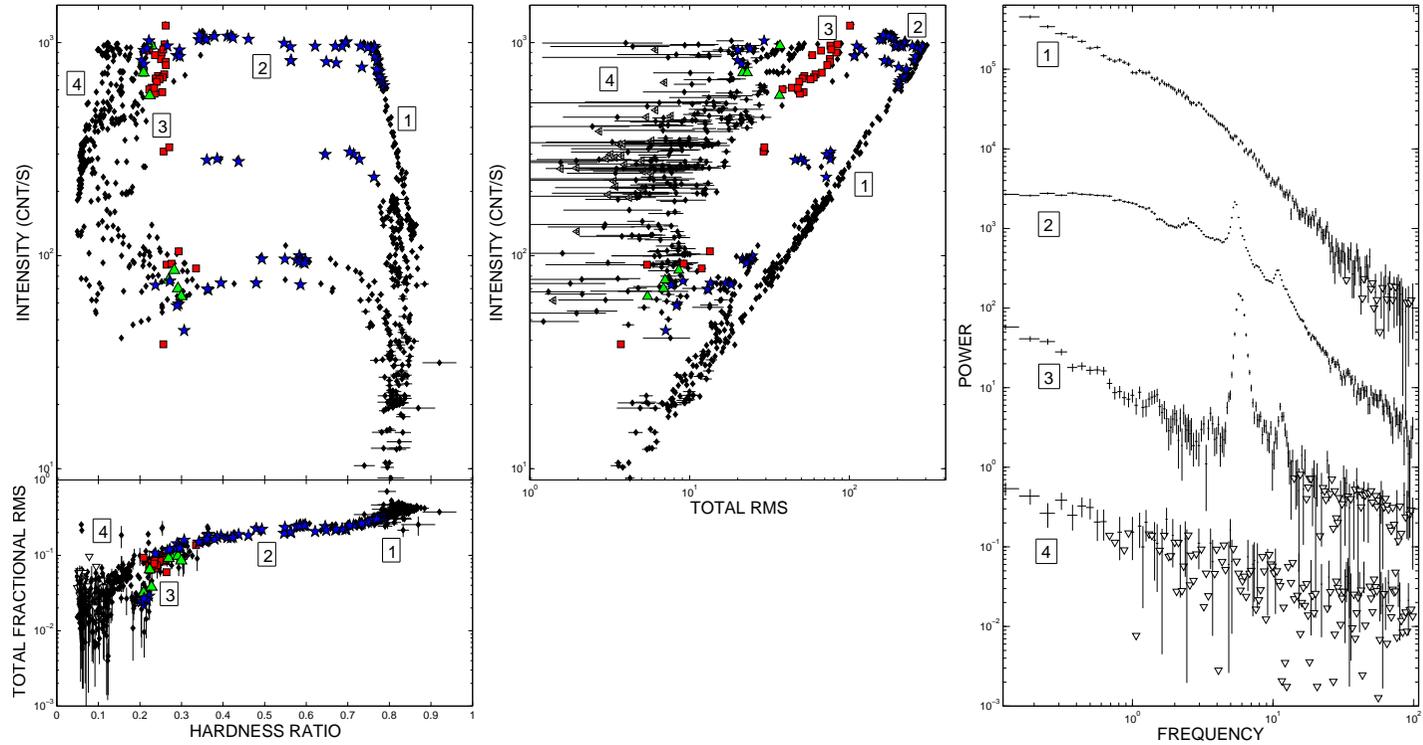}}
     \caption{Top-left panel: HID for the 2002, 2004, 2007 and 2010 outbursts of GX 339--4, with each point 
representing a single RXTE observation. 
Bottom-left panel: the corresponding rms-hardness diagram, with empty triangles representing upper limits.
The hardness ratio is defined as the ratio of net counts in the 6.1$-$20 keV
band over those in the 3.3$-$6.1 keV band.
Middle panel:  RID for the 2002, 2004, 2007 and 2010 outbursts of GX 339--4. Symbols are as follows.
Blue stars: Type-C QPOs; red squares: Type-B QPOs; green triangles: 
Type-A QPOs; black dots: all the other RXTE observations of GX 339--4 that do not show low-frequency QPOs. 
Right panel: Power density spectra for the four regions marked in the left and middle panels. The left and
middle panels have been adapted from Motta et al. (2011). \label{fig:diagrams}}
\end{sidewaysfigure}

The behaviour of BHTs is known to be easily characterized in terms of spectral states and transitions between
them. These states, dubbed ``canonical'' by \citet{1992ApJ...391L..21M}, are defined on the basis of the
spectral and timing properties displayed by the sources during their outburst phases (see Sec.\ \ref{sec:states}). 
The canonical states become apparent in a hardness--intensity diagram (HID), where the total count rate is plotted as a function of the spectral hardness (defined as the ratio of observed counts in two energy bands). Here the spectral evolution of many BHTs can be followed along a q-shaped pattern that is traced anti-clockwise. 
In the HID (see Fig.\ \ref{fig:diagrams}, top-left panel) four different branches can be identified,
corresponding to the four sides of the ``q'' \citep[the fact that the transition from hard to soft takes
place at higher flux than the reverse transition is seen in all systems, see][]{2003MNRAS.338..189M}. The
right and left vertical branches respectively match the low hard state and the high soft state already known
before the RXTE era. In between these two well-established states, the situation is rather complex and the
intermediate points that fill this region roughly correspond to those previously know as very-high state and
intermediate state. These two states show very similar properties in the spectral domain, making it
practically impossible to distinguish between them \citep[see][]{2001ApJS..132..377H}. However, when the
timing properties are taken into account, the ``intermediate region'' of the HID can  be easily divided into
two well separated areas, as they show completely different features in the time-domain (see Sec.\ \ref{sec:states}).
It is important to remark that not all transients show a regular evolution like the one which will be described here. However, the basic states and their properties can be defined for all systems. For an overview of a large number of RXTE transients see \citet{2010MNRAS.403...61D}.

The first step towards a clear distinction of the intermediate regions is made by looking at the hardness--rms
diagram (HRD, see Fig.\ \ref{fig:diagrams}, bottom-left panel), where the fractional rms, integrated over a
broad range of frequencies, is plotted versus hardness. In this diagram, unlike in the HID, most of the
points follow a correlation over a single line (no hysteresis here), extended from the softest points, where
the points show a low level of variability ($<$5\%), to the hardest points, where the variability is much higher (also exceeding 30\%).
In the region at intermediate hardness, the correlation is rather tight; however in a narrow hardness ratio band 
several points show a lower variability than the points on the main correlation (see Fig.\ \ref{fig:diagrams}). 
This behaviour already suggests the presence of two separate states in this region, corresponding to the
points on the main correlation (spanning a very large range in hardness and rms) and to those deviating 
from it (found in a narrow hardness and rms range). However, it is only by examining the fast aperiodic 
variability that the canonical states can be distinguished clearly from one another. 

Examining the Power Density Spectra (PDS) allows one to probe the fast timing properties of a source. 
It is worth mentioning that although important information can be extracted from higher-order timing 
tools \citep[such as phase/time lags, coherence and bicoherence, see][]{1996MNRAS.280..227N,2005MNRAS.357...12M}, 
these are not essential for the basic determination of states and state transitions. A number of different PDS 
shapes can be observed in BHTs, but it is now clear that we can classify them as belonging to a few basic types 
that are closely related to the position on the HID and HRD \citep[in addition there are a few more complex PDS shapes, not shown here, associated to the ``anomalous'' state found at very high accretion rate in some sources,][]{2010LNP...794...53B}. The most prominent features in the PDS  are narrow peaks, known as low frequency quasi periodic oscillations (QPOs). QPOs have been discovered in many systems and, even though their physical origin is still under debate \citep[see][and references therein]{2011arXiv1108.0540M}, they are thought to originate in the innermost regions of the accretion flows around stellar-mass black holes.

Low-frequency QPOs (LFQPOs) with frequencies ranging from a few mHz to $\sim$20 Hz  were already found in
several sources with Ginga and divided into different classes \citep[see
e.g.,][]{1993ApJ...403L..39M,1997ApJ...489..272T}. Observations performed with the RXTE have led to an
extraordinary progress in our knowledge on properties of the variability in BHBs and it was only after RXTE
was launched that LFQPOs were detected in most observed BHBs \citep[see][for recent
reviews]{2006csxs.book...39V,2006ARA&A..44...49R,2010LNP...794...53B}. Three main types of LFQPOs, dubbed
types A, B, and C, originally identified in the PDS of XTE J1550--564 \citep[see][]{1999ApJ...526L..33W,2001ApJS..132..377H,2002ApJ...564..962R}, have been seen and identified in several sources.
All the basic types of PDS (with the exception of the additional ones related to the ``anomalous'' state)
were observed in the outbursts of GX 339--4 between 2002 and 2011(see Fig.\ \ref{fig:diagrams}, right panel).

\begin{itemize}

\item The hard points in the HID (region [1] in Fig.\ \ref{fig:diagrams}, left-upper panel) show a PDS
similar to [1] in Fig.\ \ref{fig:diagrams} (right panel). Its shape can be fitted with a small number of very broad Lorentzian components. In some cases also a low-frequency QPO peak is observable at very low frequencies ($\nu \le$ 0.01 Hz). The characteristic frequencies \citep[see][]{2002ApJ...572..392B} of all these components increase with source flux, while the energy spectrum (see below) softens gradually.

\item PDS [2] in Fig.\ \ref{fig:diagrams} can be considered a high-frequency extension of PDS [1].  It is
found at intermediate hardness values (region [2] in Fig.\ \ref{fig:diagrams}, top-left panel).  The most
prominent feature is a QPO with centroid frequency varying between $\sim$0.01 and $\sim$20 Hz and a quality
factor $Q$ around 10. The characteristic frequencies of all  Lorentzian components vary together, including
that of the QPO. As in the previous case, they are correlated with hardness: softer spectra correspond to 
higher frequencies and also to lower integrated rms variability (see Fig.\ \ref{fig:diagrams}, middle panel).
The LFQPO observed here is termed ``type-C'' QPO \citep[stars in Fig.\ \ref{fig:diagrams}, see][for a precise definition of the three QPO types]{2005ApJ...629..403C,2011arXiv1108.0540M} and it always appears together with moderately strong (10--30\% rms) band-limited noise. It is often accompanied by two peaks harmonically related: one at half the frequency and one at twice the frequency, with the higher one having a similar Q to that of the fundamental; higher harmonics are often observed \citep[][]{2010ApJ...714.1065R}.
The frequency of the type-C QPO is strongly correlated with the characteristic (break) frequency of the main underlying broad-band noise components \citep[see][]{1999ApJ...514..939W,2002ApJ...572..392B}, a correlation which also extends to neutron-star binaries.

\item Over a narrow range of intermediate hardness, PDS similar to region [3] in Fig.\ \ref{fig:diagrams} can
be found (region [3] in the HID, see Fig.\ \ref{fig:diagrams}, top-left panel). These PDS feature a QPO
called ``type-B'' (squares in Fig.\ \ref{fig:diagrams}). This oscillation shows different properties from type-C QPOs. 
The fact that type-B QPOs are not simply an evolution of type-C QPOs is demonstrated by the fast transitions observed between them \citep[see][]{1997ApJ...489..272T,2003A&A...412..235N,2004A&A...426..587C,2005A&A...440..207B}. Type-B QPOs are associated with a weak power-law noise that replaces the broad Lorentzian components of the previous PDS types. Type-B QPOs show a harmonic structure similar to that of type-C QPOs (harmonic peaks and a peak at 1/2 the frequency of the main peak). While type-C QPOs span large range in frequency, type-B QPOs are limited to the range 1--6 Hz, but detections during high-flux intervals are concentrated in the narrow 4--6 Hz range \citep[see][]{2011arXiv1108.0540M}. The centroid frequency appears positively correlated with source intensity rather than hardness.

\item At hardness values systematically slightly lower than those of PDS [3], PDS showing ``type-A'' QPO can
be observed (triangles in Fig.\ \ref{fig:diagrams}). Being a much weaker and broader feature, we know less details about this oscillation. Sometimes it is only detected by averaging observations. Its frequency is always in the very narrow range 6--8 Hz and it is associated to an even lower level of power-law noise than type-B. 

\item The three types of QPOs can be separated by plotting them against the integrated fractional rms of the PDS in which they appear 
\citep[see][]{2005ApJ...629..403C,2011MNRAS.410..679M,2011arXiv1108.0540M}.

\item The softest points in the HID (region [4] in Fig.\ \ref{fig:diagrams}, left panels) show PDS ([4] in
Fig.\ \ref{fig:diagrams}, right panel) with a weak steep component, which often needs a long integration time for a detection. Weak QPOs at frequencies higher than 10 Hz are sometimes observed, as well as a steepening/break at high frequencies. The total fractional rms can be as low as 1\%.

\item In addition to the PDS shapes described, there are two types of PDS which are not always observed in BHTs.  One PDS has a featureless curved shape, the other with additional broad and narrow features \citep[see e.g.,][]{2010LNP...794...53B}. 

\end{itemize}

By the introduction of a third fundamental diagram, the rms-Intensity diagram \citep[RID, see][]{2011MNRAS.410..679M}, it becomes possible to clearly separate the canonical states without the intervention of any spectral information. To produce the HID and the RHD three variables are needed: spectral hardness, intensity (or count rate) and integrated fractional rms. The RID is  the third possible plot between these variables: the integrated rms (not fractional) as a function of the total count rate.
In the RID, the fast variability can be used as a good tracer of different accretion regimes in black hole binaries and it becomes apparent that  apart from the linear rms-flux relation (a.k.a. hard line) found during LHS \citep[see][]{2004A&A...414.1091G}, different relations are followed during the soft and intermediate states, to the extent that it is possible to identify the states from the RID alone. State transitions produce marked changes in the rms-flux relation. It is also important to notice that, in the RID, type-B QPOs (that are the defining feature of the SIMS, see Sec. \ref{sec:states}) appear in a well-defined region between 5 and 10\% fractional rms. This fact virtually removes the necessity of a detailed analysis of the PDS to separate the intermediate states (see Sec. \ref{sec:states}).
A detailed description of the RID and its properties and peculiarities can be found in \citet{2011MNRAS.410..679M}.

It is important to remark that the diagrams described above are instrument dependent and source dependent, which means that a quantitative comparison between them is not always possible. However, they are also model independent and provide very precise measurements. For a discussion of similar diagrams based on spectral fits, see \citet{2010MNRAS.403...61D}.


\section{Timing-spectral states}\label{sec:states}

The diagrams presented in the previous section lead to the definition of  small number of states, whose boundaries are rather precise and defined by the data \citep[see also][]{2005Ap&SS.300..107H,2006MNRAS.369..305B,2010LNP...794...53B}. It is important to stress that it is possible to characterize these states only by considering both spectral and timing properties, so that the concept of ``spectral'' states carries little meaning. While the states do not have direct physical connection, their definition is a necessary step in order not to miss major physical changes in the accretion flow.

\begin{itemize}

\item {\bf The Low-Hard State (LHS)}. 
Only (and possibly always, but there are quite a few cases of missed starting LHS) at the beginning and at
the end of an outburst. It corresponds to the right nearly-vertical branch in the HID (the spectrum softens
as the flux increases). The energy spectrum is hard and usually associated to thermal Comptonization
\citep{2010LNP...794...17G}. Strong ($>$ 30\% fractional rms) variability in the form of a band-limited noise
is present (see PDS [1] in Fig.\ \ref{fig:diagrams}, right panel). As flux increases, the total fractional
variability in most cases decreases. At high rates, the end of this state is marked by the leaving of the
hard line (see Fig.\ \ref{fig:diagrams}, middle panel), while the actual shape of the power density spectrum changes more gradually across the transition.

\item {\bf The Hard-Intermediate State (HIMS)}.
It corresponds to horizontal branches in the HID. It takes place after the initial LHS and before the final
LHS, as well as in the central parts of the outburst, through mini state transitions. The energy spectrum is
softer than in the LHS and at low hardness an accretion disc component starts contributing to the RXTE/PCA
band. The variability is dominated by a weaker band-limited noise and a type-C QPO (stars in Fig.\ \ref{fig:diagrams}, left and middle panels, PDS [2], right panel; for a discussion of the different types of QPO see \citealt{2005ApJ...629..403C,2011arXiv1108.0540M}).
The fractional rms decreases with hardness down to $\sim$10\%; overall, the evolution of the power spectra is such that they are an extension of the LHS ones.
The QPO frequency is anti-correlated with hardness and correlated with the photon index of the hard spectral component.

\item {\bf The Soft-Intermediate State (SIMS)}.
When both hardness and fractional rms drop below well defined thresholds (0.71 for GX 339--4, see Fig.\ \ref{fig:diagrams}), the source enter the SIMS. Here the energy spectrum is similar to that of the nearby HIMS (at least below $\sim$ 10 keV, see below), but the timing properties are completely different: the band-limited noise disappears (drop in the HRD, rms below 10\%) and is replaced by a power-law noise, while a marked type-B QPO is present (squares in 
Fig.\ \ref{fig:diagrams}, PDS [3]), whose frequency correlates with the hard X-ray flux \citep[see][]{2011arXiv1108.0540M}.

\item {\bf The High-Soft State (HSS)}. The leftmost branch in the HID. Here the spectrum is soft, dominated
by a thermal disc contribution. A hard component is observed with varying intensity. Very little variability
is observed (PDS [4] in Fig.\ \ref{fig:diagrams}, right panel), sometimes with type-A QPOs. This state spans a larger range in luminosity.

\end{itemize}

The main time evolution of a transient are: LHS -- HIMS -- SIMS - HSS as accretion rate increases, followed by HSS -- SIMS -- HIMS -- LHS as accretion rate decreases \citep[for a discussion of the reverse transition from soft to hard see][]{2006ApJ...639..340K,2006MNRAS.369..305B}. Intermediate minor transitions between states, usually not involving the LHS, can be observed. In some sources the initial LHS (or LHS -- HIMS) is so fast that it is not observed. Some sources never leave the LHS, in very few cases reaching the HIMS without showing a transition to the SIMS \citep[][]{2009MNRAS.398.1194C,2010MNRAS.408.1796M}. Very few sources show a more complex time evolution, which however involves observations that can be classified in one of the four states outlined above. In addition, an ``anomalous state'' is present in some sources, related to very high luminosities \citep[see][]{2010LNP...794...53B,2006MNRAS.371.1216D}.
The level at which the LHS -- HIMS transition takes place can vary in one source from outburst to outburst, but it is always higher than the reverse transition and their level appear to be anti-correlated: a brighter upper HIMS branch corresponds to a fainter lower HIMS branch \citep[see][]{2003A&A...409..697M,2010MNRAS.403...61D}. What physical quantity determines the flux (and accretion rate) level at which the hard to soft transition takes place is currently unknown, although it has been suggested that the time from the previous outburst plays a role \citep[][and references therein]{2009ApJ...701.1940Y}.

\section{Energy spectra}

Timing properties are crucial to define the states, but although few models have been proposed to explain the origin and the nature of time-variability, still there is not a clear interpretation nor an explanation on their physical origin. 

Even though we are still far from achieving a clear and complete comprehension of the full accretion
mechanism acting in X-ray binaries, we have an idea of the main physical processes involved. Here we will
discuss spectral variations along the HID and we will mainly focus on the changes that take places during
spectral transitions. Note that the information we discuss in this section is mainly based on the 2--200 keV
energy band since a good portion of it comes from RXTE data. Very detailed spectra come from instruments with
higher spectral resolution. 

The emission coming from X-ray binaries cover a very broad energy interval, extending from the radio to the very high energies (hard X-rays and sometimes gamma-rays). The main effects of the accretion mechanism in the region close to the compact object can be observed in the X-rays, but other fundamental processes are responsible for the emission at very different wavelengths (see also Sec. \ref{sec:multi}). 
The X-rays consist of both thermal emission coming from the very inner regions of an optically thick accretion disc as well as a harder component, extending to hundreds of keV and sometimes up to an MeV. The characteristics of this hard component are state-dependent. In the hard states (LHS and HIMS), it is usually interpreted as thermal or hybrid Compton radiation from the interaction of the soft disc photons with relativistic electrons in the inner regions of the accretion flow \citep[see][for recent reviews]{2007A&ARv..15....1D,2010LNP...794...17G}. It has also been suggested that non-thermal Comptonization processes (e.g., self-Compton from synchrotron emission from the jet or bulk motion Comptonization) could be responsible for this component \citep{2001ApJ...562L..67L,2010LNP...794..143M}. 
For the soft states (HIMS and HSS), the hard component extends to higher energies and is possibly of non-thermal origin.

\begin{figure}
     \centerline{\includegraphics[width=13.5cm]{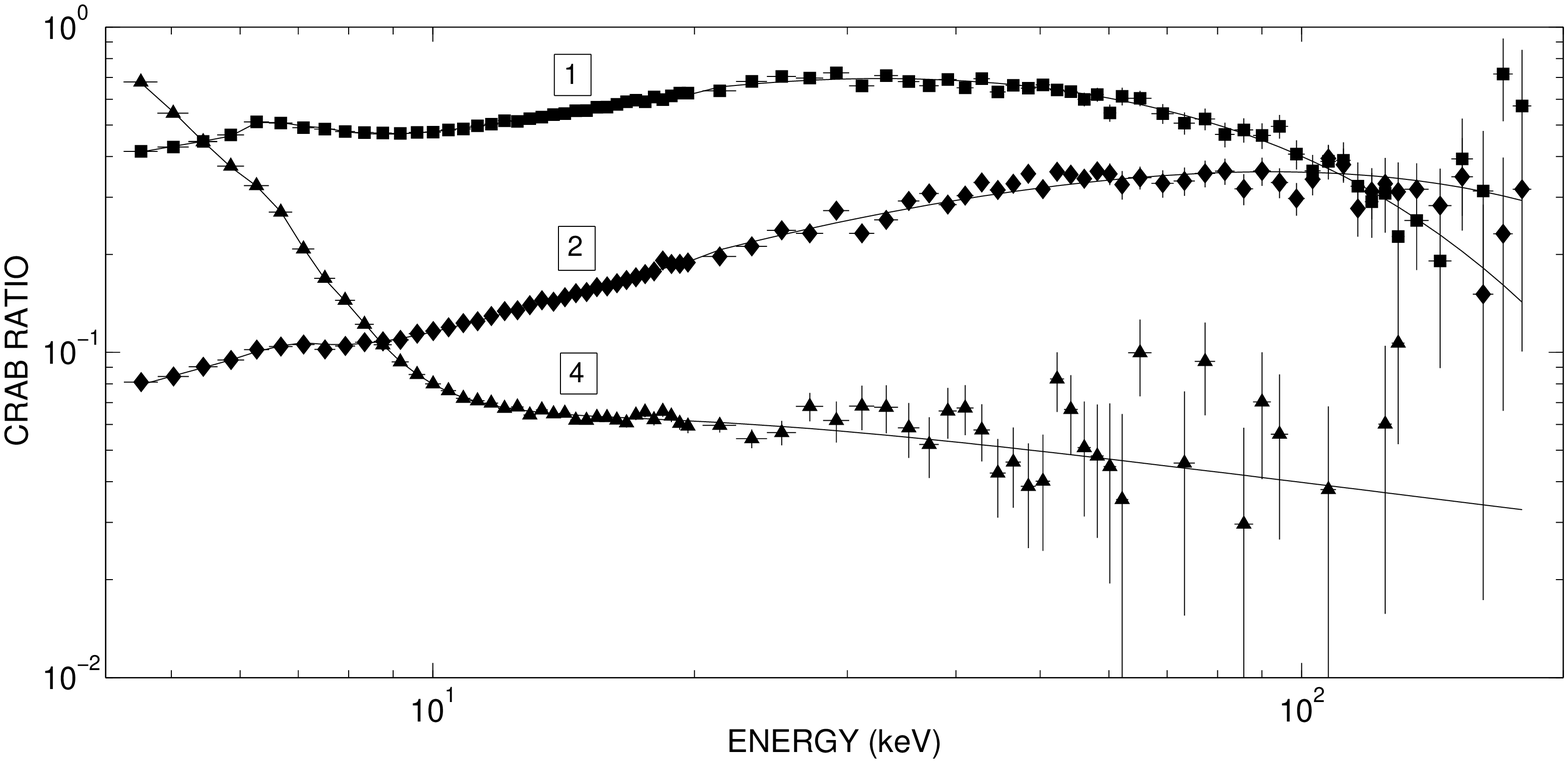}}
     \caption{PHA ratio of GX 339--4 to Crab for three selected observations. (1) LHS spectrum, (2) HIMS
spectrum, (4) HSS spectrum. Figure adapted from \citet{2009MNRAS.400.1603M}.} \label{fig:spectra}
\end{figure}

The movement of a source along the HID must be decomposed in vertical (i.e. count rate) and horizontal (i.e. spectral hardness) excursions. The vertical movement, typical of the hard state where the spectrum softens less dramatically, is thought to be driven by accretion rate, depending on the geometry of the system and of the accretion flow. The horizontal movement is caused by a combination of a steepening of the hard component and the appearance of a thermal disc component in the RXTE band. While accretion rate most likely continues increasing, the marked change in accretion properties is triggered by a hitherto unknown parameter \citep[but see][]{2009ApJ...701.1940Y}. 

\begin{figure}
     \centerline{\includegraphics[width=13.5cm]{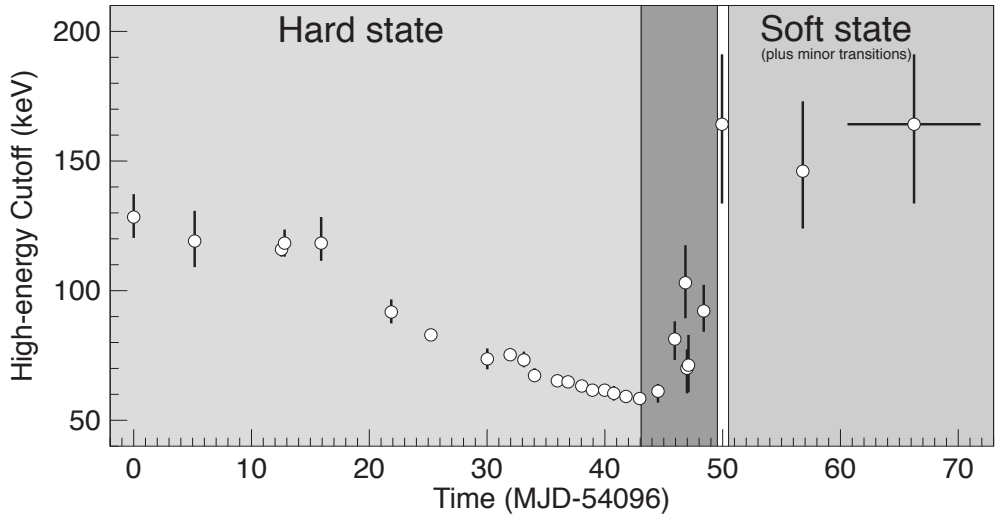}}
     \caption{Evolution of the high-energy cutoff in the 2007 outburst of the BHT GX 339--4.} \label{fig:cutoff}
\end{figure}

\begin{itemize}

\item {\bf LHS} ([1] in Fig.\ \ref{fig:diagrams}) -- as already mentioned, all outbursts most likely start in
this state, although at times it is not observed, probably because of its short time scale. The corresponding
broad-band energy spectra are very hard and usually show a high-energy cutoff around 50--100 keV (spectrum 1
in Fig.\ \ref{fig:spectra}). 
In GX 339--4 and GRO J1655--40 this cutoff was seen to move to lower energies as flux increased (see Fig.\ \ref{fig:cutoff}), together with a steepening of the photon index.
A faint cold disc with a large inner radius and a peak temperature of few tenths of keV  is sometimes observable if the interstellar absorption is low \citep[see e.g.,][]{2010MNRAS.404L..94M}. However, in few sources a hot thermal disc has been detected, suggesting that in some cases the inner disc radius could be always close to the innermost stable orbit. 
Sometimes an additional non-thermal component is also present in the spectrum \citep[][]{2002APS..APRN17105M}. This component  has been proposed to be associated with the synchrotron jet detected at radio waves. 

\item {\bf HSS}  ([4] in Fig.\ \ref{fig:diagrams}) --  the energy spectrum is dominated by the thermal
emission from the disc, observed with a small inner disc radius and a temperature that reaches typical values
of $\sim$1 keV. The hard component, probably the result of Comptonization of soft photons on a non-thermal
distribution of relativistic electrons, is in general very faint and variable without an observed cutoff
below 1 MeV (spectrum \#4 in Fig.\ \ref{fig:spectra}). 

\item {\bf HIMS/SIMS}  ([2] and [3] in Fig.\ \ref{fig:diagrams}) -- during the HIMS and the SIMS, the
transition from hard to soft state takes place. The spectrum is a combination of the soft and hard component
dominating the HSS and LHS respectively (spectrum 2 in Fig.\ \ref{fig:spectra}). In several sources the inner
disk radius has been observed to move inward during these states. This behaviour provides a physical
explanation to the changes in the spectral shape during the transition, partially produced by the increasing
soft emission coming from the disc. Whether the cutoff decreases, increases or disappears in the intermediate states has been a topic of debate for several years. This has been partially solved thanks to the good coverage achieved for some sources. 
In GX 339--4 \citep[][see Fig.\ \ref{fig:cutoff}]{2009MNRAS.400.1603M} and GRO J1655--40 \citep[][]{2008ApJ...679..655J}, after the decrease during the LHS, the high-energy cutoff is observed to suddenly increase in the HIMS, to reach its maximum in the SIMS. If a high energy cutoff is present in the HSS, it is not observable below 1 MeV \citep[][]{2008MNRAS.390..227D}. The decrease seen during the LHS is thought to be produced by the cooling of the Comptonizing medium due to the higher fraction of soft photons undergoing Comptonization, while the reason for the HIMS increase is still under discussion.
Spectra from the softest HIMS and from the SIMS are very similar, since their hardness is very similar and the evolution from one to the other is smooth \citep[for a discussion, see][]{2011arXiv1108.2198S}.

\end{itemize}

As mentioned above, some sources (e.g., GRO 1655--40) display another state, the ``anomalous'' or ``ultra-luminous'' state. Here, the source shows very high luminosities and the energy spectrum appears dominated by a soft disc component. The disc component allows to infer a high inner-disc temperature associated to a small inner disc radius, while the hard component is steep and faint.

\section{Rms spectra}

The amount of variability as a function of energy is another observable which is seen to vary along the outburst. This is expected, since the relative contribution of each spectral component strongly depends on the accretion state. Apart from its average spectral shape, usually measured over long time scales ($\geq$ 1 ks), each component also prompts a certain level of variability in the observed light-curve. The shape of the rms spectrum (i.e., fractional rms as a function of energy) encodes information directly related to the relative contribution of each emission process and the intrinsic origin of the variability itself. Whereas the former can also be obtained from spectral studies, the latter results in unique insights into accretion physics.

\begin{figure}
     \centerline{\includegraphics[width=14.0cm]{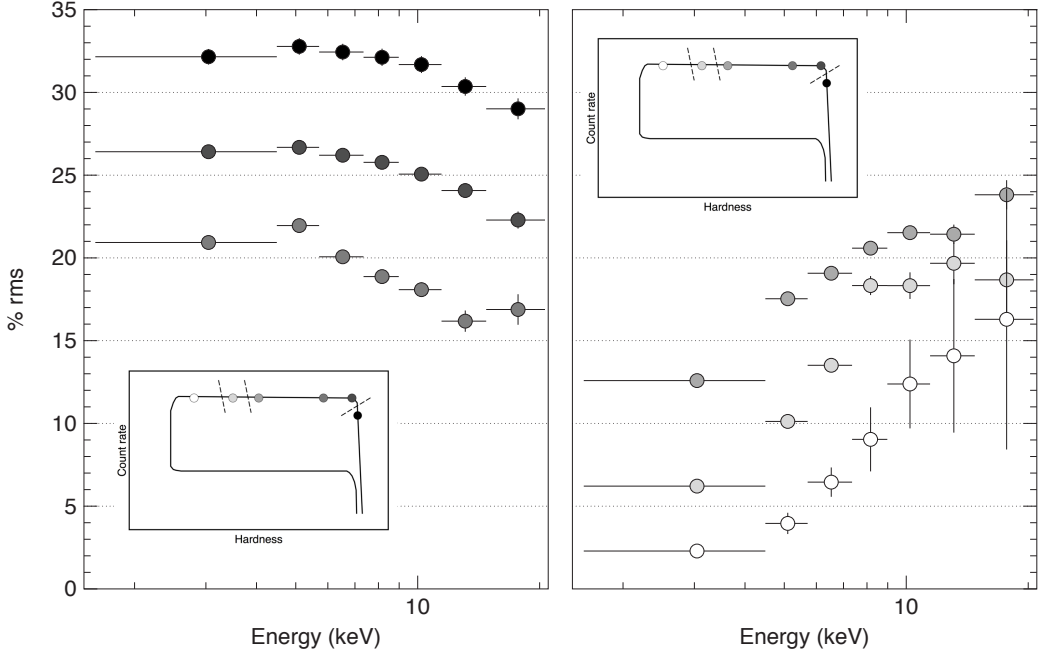}}
     \caption{Evolution of the 0.1--32 Hz rms spectrum along the hard-to-soft transition of GX 339--4. 
     Left panel, from top to bottom: rms spectra from a bright LHS, a hard HIMS and a softer HIMS (see inset).
     Right panel, from top to bottom: rms spectra from a HIMS just before the transition, a SIMS and a HSS (see inset).
     \label{fig:rmss}}
\end{figure}

As for many other cases, the RXTE/PCA has made a strong contribution, most of the results present in the
literature being referred to the $\sim$ 2--20 keV energy band. In a first approximation, rms spectra with
three different shapes are observed in BHT: flat, rms decreasing with energy (soft; aka., inverted) and
increasing with energy (hard). The first two cases correspond to the hard states (see below), whereas a hard
rms spectrum is seen as the system softens. In Fig.\ \ref{fig:rmss} we show an example of these shapes using RXTE/PCA data of GX 339-4 during its hard-to-soft transition.   

\begin{itemize}
\item \textbf{LHS}: Flat and inverted rms spectra are observed during this state. Clear examples of the
former are e.g., XTE J1550--564 (\citealt{2005MNRAS.363.1349G}), Cyg X--1 or XTE 1752--223
(\citealt{2010MNRAS.404L..94M}). The latter behaviour is observed in XTE J1650--500
(\citealt{2005MNRAS.363.1349G}) or Swift 1753.5--0127 (Soleri et al., submitted). Flat rms spectra are easily
explained by considering that variability arises from variations in the normalization of one single spectral
component. This could account for what we see in the hard state, where the hard component is dominant ($\geq
2$ keV; see Sect. 4). Variations in the spectral shape of this component (i.e., not only in normalization)
need to be taken into account to explain inverted rms spectra. \cite{2005MNRAS.363.1349G} (see also
\citealt{2002ApJ...578..357Z}) show that a strongly varying (disc) seed photon input can produce inverted rms
spectra, since it varies the spectral shape by making the power-law to pivot around $\sim 50$ keV. Variations
in the normalization of the power-law component are also needed to reproduce the observed dependence of the
rms with energy. Interestingly, in some of the cases where flat rms spectra are observed, they are still
consistent with being inverted at $\sim$ 1--2\% level (e.g., Fig.\ \ref{fig:rmss}).

\item \textbf{HIMS:} As it happens for several spectral and timing parameters, a major change in the shape of
the rms spectra is observed along the HIMS. Whereas LHS-like rms spectra are observed during the early stages
of the HIMS, they harden as the system approaches the soft states. This transition occurs fast; for instance,
the two last HIMS spectra presented in Fig.\ \ref{fig:rmss} are separated by five days. Their corresponding energy spectra are different as well, since the disc component is only present in the second one. \cite{2011MNRAS.410..679M} observed a flux increase at constant absolute variability in correspondence to the appearance of the disc component in the energy spectra of this system during the HIMS. The above suggests disc variability to be low at least during last (softest) stages of the HIMS.  

\item \textbf{SIMS and HSS:} Hard rms spectra are observed during soft states. As shown in 
Fig.\ \ref{fig:rmss}, these are remarkably similar to those observed at the softest end of the HIMS, but with
total variability decreasing as the system softens. These patterns can be explained by considering no or very
little disc variability, which results in lower rms values in the soft band when disc dominates. Above 5--10 keV, where the variable power-law dominates, we recover the flat shape observed in the LHS.  

\end{itemize} 

\subsection{The origin of the variability and the role of the disc component}
It has been known for years that variability is mainly associated with the hard component rather than the
disc. During soft states, where the disc dominates, low variability levels ($\leq$ 5 per cent) are observed.
Even there, its energy dependence also points to a hard-component origin (see e.g.,
\citealt{2010LNP...794...17G} and references therein). HIMS observations also point in the same direction (see above) and in the LHS, where only the hard component is present above $\sim 2$ keV, we see the highest variability levels. Paradoxically, Comptonization models aiming at explaining the inverted rms spectra require a highly variable disc during the LHS. Softer ($\leq 2$ keV) XMM observations have shed some light on this issue. Evidence for low frequency ($ \lesssim$1 Hz) disc variability during the LHS has been reported by \citet{2009MNRAS.397..666W}. Also at low frequencies, \citet{2011MNRAS.414L..60U} published time-lag measurements showing disc variability leading hard-component variability. Both results represent a strong evidence for a variable disc component during the LHS at low frequencies \citep{2005MNRAS.363.1349G}, whereas for time scales shorter than $\sim 1$ s (i.e. those typically used for variability studies) results are not conclusive. 
        
If variability originates in the soft component, it should progressively disappear to explain the lack of variability observed as the energy spectrum softens. Hard-component variability should decrease as well, but high levels of hard-component variability are observed during soft states. Interestingly, \citet{2011MNRAS.415..292M} report a dramatic fade of hard-component variability during the SIMS in MAXI J1659--152. The light curve associated to the disc component is also shown to be more stable at very low frequencies once the system reaches the HSS.\\
We note that the role played by the jet component, which is probably highly variable (e.g.,
\citealt{2010MNRAS.404L..21C}) and has been proposed to account in some cases for the observed X-ray emission
during the LHS (e.g., \citealt{2001A&A...372L..25M}) should be much better understood to solve this problem.
The combination of multiwavelength studies and the broader spectral coverage planned for future X-ray
missions, such as LOFT (\citealt{2011arXiv1107.0436F}) and ASTROSAT \citep[][]{2006AdSpR..38.2989A}, 
should make a decisive contribution to this issue.

\section{High-frequency oscillations}

Despite the large number of available RXTE observations of black hole binaries, only a handful of detections of quasi-periodic peaks above 30 Hz is available \citep[see][and references therein]{2006MNRAS.369..305B}. Three sources show a single QPO peak, although for two of them the detection is not very significant and in the case of XTE J1650--500 the peak appears only after stacking spectra from a specific state. Four other sources show two peaks, which when detected more than once appear at almost the same frequency: GRO J1655--40 (300 and 450 Hz), XTE J1550--564 (184 and 276 Hz), H 1743--322 (165 and 241 Hz) and GRS 1915+105 (41 and 69 Hz). In the first three cases, the frequencies are consistent with being in 3:2 ratio, for GRS 1915+105 the ratio is 5:3.
Interestingly, an additional peak at 27 Hz was found in GRS 1915+105, which is in 2:3 ratio with 41 Hz \citep[][]{2001A&A...372..551B}.
These features are very important as they constitute the highest-frequency signals from accreting black holes and they are most likely related to effects of General Relativity in the strong field regime.
This is not the place to discuss more details or applicability of models, but it is important to remark that all these detections correspond to sources in (or close to) the SIMS. The case of GRS 1915+105 is more complex, as the association of its features with the four states described above is not simple \citep[see][]{2003A&A...412..229R,2008MNRAS.383.1089S}.

\section{The multi-wavelength view}\label{sec:multi}

The phenomenology described above was found to have strong links with the properties of relativistic jets as observed in the infrared and radio bands. 
The radio emission offers the possibility to study the synchrotron emission related to the relativistic jets \citep[see][]{2010LNP...794..115F,2010LNP...794...85G}, while in the IR, optical and UV bands the thermal emission from the companion star and from the outer edge of the disc can be observed \citep[see e.g.,][]{2005ApJ...624..295H,2006MNRAS.371.1334R,2007ApJ...670..610M}.

\begin{figure}
     \centerline{\includegraphics[width=15.0cm]{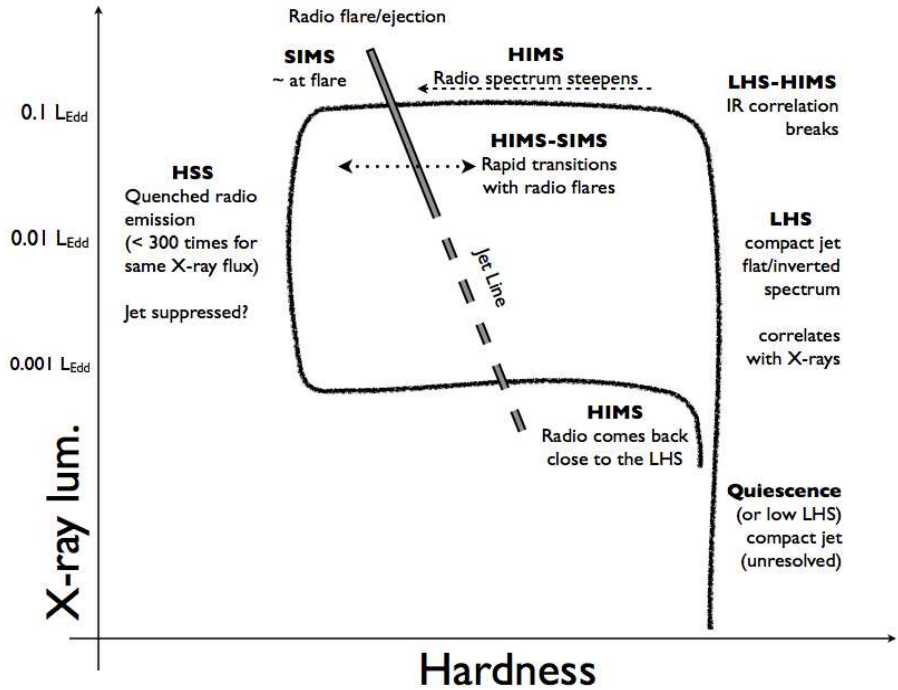}}
     \vspace{-1cm}
     \caption{Stylized hardness-intensity diagram for a BHT with labels indicating the radio properties. \label{fig:radio}}
\end{figure}

The basic properties are summarized in Fig.\ \ref{fig:radio}.
In the LHS, a steady compact jet is observed, with a flat or inverted spectrum and a barely resolved spatial distribution. The radio and infrared fluxes are correlated with the X-ray flux, correlation which extends to the AGN when the proper mass scaling is considered \citep[see][and references therein]{2011MNRAS.414..677C}. The infrared flux also correlates with the X-ray one.

At the top of the LHS branch, in GX 339--4 the correlation between X-ray and infrared was observed to break
down \citep[][]{2005ApJ...624..295H}. As the source follows the HIMS and approaches the transition to the
SIMS, the radio spectrum starts steepening \citep{2004MNRAS.355.1105F}. Around, but not exactly at the time
of the transition
(\citealt{2009MNRAS.396.1370F}), a radio flare is observed (or the ejection time of a fast relativistic resolved jet). The hardness threshold at which this happens has been dubbed ``jet line''. After the source enters the HSS, no nuclear radio emission has been detected up to know, as all detections are compatible with emission from the ejecta. Recent observations of 4U 1957+11 have put the limit to more than 300 times lower than the LHS radio flux at the same X-ray flux \citep[][]{2011arXiv1106.0723R}.

In correspondence of additional transitions back and forth between HIMS and SIMS, there is some evidence that more radio flares are emitted in XTE J1859+226, \cite[see][]{2002MNRAS.331..765B}, and in GRS 1915+105 where the oscillations have been connected to such transitions \cite[][]{2004ARA&A..42..317F}.

On the return transition, the radio emission becomes detectable again, but at a higher hardness value than it disappeared earlier, when the source has already reached the LHS \citep[][]{2006ApJ...639..340K}. The radio/X-ray correlation is re-established and can be followed down to very low accretion rate \citep[][]{2006MNRAS.370.1351G}.

\subsection{Optical and infrared timing}
The advent of new instrumentation during the last decade (e.g., ULTRACAM; \citealt{2007MNRAS.378..825D}) has
enabled us to extend some of the timing studies, classically performed at high energies, to the optical and
infrared domains with unprecedented quality. In contrast to what is typically observed in neutron stars
(e.g., \citealt{2007MNRAS.379.1637M}), complex cross correlation functions, with optical variability leading
and anti-correlated to X-rays were found in BHT during the LHS (e.g., \citealt{2001Natur.414..180K};
\citealt{2008ApJ...682L..45D}). These features cannot be explained solely by reprocessing on the accretion
disc (see \citealt{2003MNRAS.345..292H}) and theoretical interpretations proposed generally require of a
substantial contribution to the optical emission from the synchrotron jet observed at radio waves
(\citealt{2004MNRAS.351..253M}; but see \citealt{2011ApJ...737L..17V}). This interpretation is also supported by some
spectral energy distribution studies, which point to a significant jet contribution to the OIR
(optical-infrared) and maybe to higher energies during the LHS (e.g., \citealt{2001A&A...372L..25M}; \citealt{2010MNRAS.405.1759R}) and by high time resolution studies performed in the near IR \citep{2010MNRAS.404L..21C}.
To gauge the jet contribution to the OIR is an important open issue since it has strong implications in, e.g., understanding the amount of energy carried by the jet. New studies are being currently undertaken by several groups from both spectral and timing point of view, and much progress is expected in the forthcoming years in this relatively young field. Here, it is worthy to notice the timing capabilities of ASTROSAT in the UV, that could be exploited in the very near future.

\section{Dynamical measurements of black hole masses}

At present, more than $\sim 20$ of X-ray binaries where the compact object is heavier than 3$M_{\odot}$ have been found, with masses as high as $\sim 15 M_{\odot}$ (\citealt{2007Natur.449..872O}).
Most of these measurements come from BHT, where the companion star is usually detectable during the quiescence epoch and classical techniques based on the Kepler laws (see e.g., \citealt{2007IAUS..238....3C} for a recent review) can be applied. However, this number is still not enough to have a good description of the mass spectrum of stellar mass BHs, and some open questions remain (e.g., is there a real gap in the distribution of masses of compact objects between NS and BHs?; see \citealt{2010ApJ...725.1918O}). On the other hand, the mass is a key parameter for BH spin measurements (e.g., \citealt{2011CQGra..28k4009M}).
New analysis methods (e.g., \citealt{2003ApJ...583L..95H}; \citealt{2008MNRAS.385.2205M}, \citealt{2010ApJ...710.1127C}) and observing facilities (e.g., \citealt{2011MNRAS.413L..15C}) are allowing us to tackle faint systems and to infer more accurate masses. A big improvement will come with the next generation of telescopes (20--40 m), which will enable to study objects with very faint quiescence levels. Still, one of the main problems will be to discover more of those systems without the need that they go into outburst. This would increase significantly the number of dynamical mass measurements since the population of quiescence black holes is expected to be at least of the order of few thousands (see e.g., \citealt{1998A&A...333..583R} and references therein).

\section{Where do we go from here: ASTROSAT and LOFT}

The natural successor of RXTE is the Indian multi-wavelength mission ASTROSAT
\citep[][]{2006AdSpR..38.2989A}, which is currently schedule for launch by the second half of 2012. While its
all-sky monitor (SSM) will allow good coverage of the transient X-ray sky and its UV telescope (UVIT) will
observe the very few BHTs which are not affected by very high absorption, its X-ray pointed instruments will
extend the RXTE characteristics and make it possible to continue and extend the study of bright BHT
transients. The soft X-ray telescope (SXT: 0.3--8 keV) and the coded-mask imager (CZTI: 10--150 keV) will
complement the large proportional counter array (LAXPC: 3--80 keV) to produce broad-band X-spectra, and the LAXPC effective area (6000 cm$^2$ at 10 keV) will yield the large photon counting necessary for fast-timing studies.

\begin{figure}
     \centerline{\includegraphics[width=13.8cm]{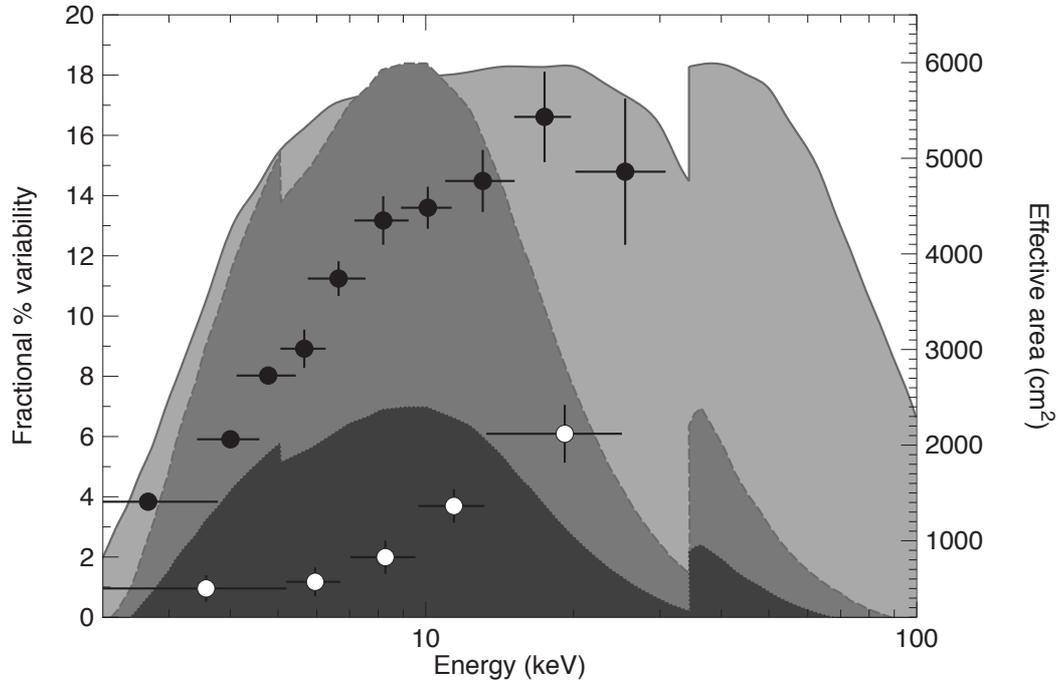}}
     \caption{A comparison of instrument effective areas and energy-dependent variability for selected signals.
     Dark gray: area of RXTE/PCA (2 units); medium gray: RXTE/PCA effective area (5 units); light gray: ASTROSAT/LAXPC effective area (3 units). The black points report the energy dependent fractional rms variability of the type-C QPO in XTE J1859+226 \citep[from][]{2004A&A...426..587C}, the white points that of the HFQPO in GRS 1915+105 \citep[from][]{1997ApJ...482..993M}.
     \label{fig:astrosat}}
\end{figure}

For the phenomenology described above, the LAXPC effective area is particularly interesting. While comparable
to that of the RXTE/PCA below 10 keV, between 10 and 80 keV the LAXPC will be much more sensitive (in
particular since later PCA observations are performed with 1 or 2 detector units, see Fig.\
\ref{fig:astrosat}). All timing signals from BHTs, with the exception of the strong noise in the LHS, are
stronger at high energies. 
In Fig.\ \ref{fig:astrosat} are shown the energy dependence of a type-C QPO and a HFQPO, superimposed on the PCA and LAXPC effective areas. It is clear that the higher sensitivity of the LAXPC above 10 keV will allow the study of weaker signals and in particular will offer the possibility to detect new HFQPOs in addition to the handful already available.

As the observational constraints of ASTROSAT will not be the same as those of RXTE, it is important to plan
the best observational strategy for BHTs. An extremely dense coverage of outbursts such as that in Fig.\ \ref{fig:diagrams} is most likely not going  to be available, at least in the first phases of the mission. However, the detailed knowledge of the phenomenology described in the previous sections make it sure that even selected pointings, triggered by other instruments such as the SSM, MAXI, INTEGRAL/IBIS or Swift/BAT, have the potential of a late yield. In particular, a strategy of fewer longer observations in correspondence of main transition and states will generate crucial data, especially if coordinated with observations at lower wavelengths (radio, IR) from the ground.

For the most distant future, not before 2020, the LOFT mission has been proposed and is being evaluated \citep[][]{2011arXiv1107.0436F}. The effective area of the LAD instrument is currently being planned to be 12 m$^2$ at 8 keV, clearly a major step forward compared to current instrumentation. With LOFT, it will be possible to detect HFQPOs at very low fractional rms, therefore unveiling its nature by examining the distribution in detected frequencies (one of the current models predicts that the frequencies are constant), while the relative variation of the frequencies will tell us to which relativistic time scales, if any, they correspond to. Moreover, a very large collecting area will allow the detection of a sufficiently high number of photons to study low-frequency oscillations directly in the time domain, opening the possibility of understanding the nature of their quasi-periodicity and its connection to accretion and General Relativity. The good energy resolution will also make it possible to link fast variations to spectral measurements, tying two parallel aspects connected to relativistic effects, a subject that it is barely possible to touch with RXTE.


\section*{Acknowledgements}

SM and TB acknowledge support from grant ASI-INAF I/009/10/. The research leading to these results has
received funding from PRIN INAF 2007 and from the European Community's Seventh Framework Programme
(FP7/2007-2013) under grant agreement number ITN 215212 `Black Hole Universe'. Partially funded by the
Spanish MEC under the Consolider-Ingenio 2010 Program grant CSD2006-00070: `First Science with the GTC'
(http://www.iac.es/consolider-ingenio-gtc/).

\label{lastpage}

\begin{thebibliography}{}

 \bibitem[Agrawal(2006)]{2006AdSpR..38.2989A} Agrawal P.~C., 2006, AdSpR, 38, 2989 

 \bibitem[Belloni(2010)]{2010LNP...794...53B} Belloni T.~M., 2010, in Belloni T., ed, 
         The jet paradigm: from  microquasars to quasars, Lecture Notes in Physics, Springer Verlag, Berlin, 794, 53
 
 \bibitem[Belloni et al.(2001)]{2001A&A...372..551B} Belloni T., M{\'e}ndez M., S{\'a}nchez-Fern{\'a}ndez C., 2001, A\&A, 372, 551
 
 \bibitem[Belloni et al.(2002)]{2002ApJ...572..392B} Belloni T., Psaltis D., van der Klis M., 2002, ApJ, 572, 392 
 
 \bibitem[Belloni et al.(2005)]{2005A&A...440..207B}  Belloni T., Homan J., Casella P., van der Klis M., Nespoli E., Lewin W.~H.~G., Miller J.~M., M{\'e}ndez M., 2005, A\&A, 440, 207
  
  \bibitem[Belloni et al.(2006)]{2006MNRAS.369..305B} Belloni T., Soleri P., Casella P., M{\'e}ndez M., Migliari S., 2006, MNRAS, 369, 305
  
  \bibitem[Brocksopp et al.(2002)]{2002MNRAS.331..765B} Brocksopp C., et al., 2002, MNRAS, 331, 765
  
  \bibitem[Cantrell et al.(2010)]{2010ApJ...710.1127C} Cantrell A.~G., et al.\ 2010, ApJ, 710, 1127 
  
  \bibitem[Capitanio et al.(2009)]{2009MNRAS.398.1194C} Capitanio F., Belloni T., Del Santo M., Ubertini P., 2009, MNRAS, 398, 1194
  
  \bibitem[Casares(2007)]{2007IAUS..238....3C} Casares J., 2007, IAU Symposium, 238, 3 
  
  \bibitem[Casares et al.(1992)]{1992Natur.355..614C} Casares J., Charles P.~A., Naylor T., 1992, Nature, 355, 614 
  
  \bibitem[Casella et al.(2004)]{2004A&A...426..587C} Casella P., Belloni T., Homan J., Stella L., 2004, A\&A, 426, 587
  
  \bibitem[Casella et al.(2005)]{2005ApJ...629..403C} Casella P., Belloni T., Stella L., 2005, ApJ, 629, 403
  
  \bibitem[Casella et al.(2010)]{2010MNRAS.404L..21C} Casella P., et al., 2010, MNRAS, 404, L21
  
   \bibitem[Coriat et al.(2011)]{2011MNRAS.414..677C} Coriat M., et al., 2011, MNRAS, 414, 677
   
   \bibitem[Corral-Santana et al.(2011)]{2011MNRAS.413L..15C} Corral-Santana J.~M., Casares J., Shahbaz T., 
         Zurita C., Mart{\'{\i}}nez-Pais I.~G., Rodr{\'{\i}}guez-Gil P., 2011, MNRAS, 413, L15 
   
   \bibitem[Del Santo et al.(2008)]{2008MNRAS.390..227D} Del Santo M., Malzac J., Jourdain E., Belloni T., Ubertini P., 2008, MNRAS, 390, 227
   
   \bibitem[Dhillon et al.(2007)]{2007MNRAS.378..825D} Dhillon V.~S., et al., 2007, MNRAS, 378, 825 
   
   \bibitem[Done \& Kubota(2006)]{2006MNRAS.371.1216D} Done C., Kubota A., 2006, MNRAS, 371, 1216

  \bibitem[Done et al.(2007)]{2007A&ARv..15....1D} Done C., Gierli{\'n}ski M.,  Kubota A., 2007, The Astronomy and Astrophysics Review, 15, 1 
  
  \bibitem[Dunn et al.(2010)]{2010MNRAS.403...61D} Dunn R.~J.~H., Fender R.~P., K{\"o}rding E.~G., Belloni T., Cabanac C., 2010, MNRAS, 403, 61
  
  \bibitem[Durant et al.(2008)]{2008ApJ...682L..45D} Durant M., Gandhi P., Shahbaz T., Fabian A.~P., Miller J., 
   Dhillon V.~S.,  Marsh, T.~R., 2008, ApJ, 682, L45 

 \bibitem[Elvis et al.(1975)]{1975Natur.257..656E} Elvis M., Page C.~G., Pounds K.~A., Ricketts M.~J., Turner
M.~J.~L., 1975, Nature, 257, 656
 
 \bibitem[Fender \& Belloni(2004)]{2004ARA&A..42..317F} Fender R., Belloni T., 2004, ARA\&A, 42, 317
 
 \bibitem[Fender et al.(2004)]{2004MNRAS.355.1105F} Fender R.~P., Belloni T.~M., Gallo E., 2004, MNRAS, 355, 1105

 \bibitem[Fender et al.(2009)]{2009MNRAS.396.1370F} Fender R.~P., Homan J., Belloni T.~M., 2009, MNRAS, 396, 1370

 \bibitem[Fender(2010)]{2010LNP...794..115F} Fender R., 2010,  in Belloni T., ed, 
       The jet paradigm: from  microquasars to quasars, Lecture Notes in Physics, Springer Verlag, Berlin, 794, 115
 
 \bibitem[Feroci et al.(2011)]{2011arXiv1107.0436F} Feroci, M., \& LOFT Consortium, 2011, arXiv:1107.0436
 
 \bibitem[Gallo(2010)]{2010LNP...794...85G}  Gallo E., 2010, in Belloni T., ed, 
         The jet paradigm: from  microquasars to quasars, Lecture Notes in Physics, Springer Verlag, Berlin, 794, 85

 \bibitem[Gallo et al.(2006)]{2006MNRAS.370.1351G} Gallo E., Fender R.~P., Miller-Jones J.~C.~A., Merloni A., Jonker P.~G., Heinz S., Maccarone T.~J., van der Klis M., 2006, MNRAS, 370, 1351
 
 
 \bibitem[Gierli{\'n}ski \& Zdziarski(2005)]{2005MNRAS.363.1349G} Gierli{\'n}ski M., Zdziarski, A.~A., 2005, MNRAS, 363, 1349
 
 \bibitem[Gilfanov(2010)]{2010LNP...794...17G} Gilfanov, M.\ 2010, in Belloni T., ed, The jet paradigm: from 
             microquasars to quasars, Lecture Notes in Physics, Springer Verlag, Berlin, 794, 17
 
 \bibitem[Gleissner et al.(2004)]{2004A&A...414.1091G} Gleissner T., Wilms J., Pottschmidt K., Uttley P., 
         Nowak M.~A., Staubert R., 2004, A\&A, 414, 1091 
 
 \bibitem[Homan et al.(2001)]{2001ApJS..132..377H} Homan J., Wijnands R., van der Klis M., Belloni T., 
      van Paradijs J., Klein-Wolt M., Fender R., M{\'e}ndez M., 2001, ApJS, 132, 377
 
 \bibitem[Homan \& Belloni(2005)]{2005Ap&SS.300..107H}  Homan J., Belloni T., 2005, Ap\&SS, 300, 107
 
 \bibitem[Homan et al.(2005)]{2005ApJ...624..295H} 
Homan J., Buxton M., Markoff S., Bailyn C.~D., Nespoli E., Belloni T., 2005, ApJ, 624, 295

\bibitem[Hynes et al.(2003)]{2003MNRAS.345..292H} Hynes R.~I., et al., 2003, MNRAS, 345, 292 

\bibitem[Hynes et al.(2003)]{2003ApJ...583L..95H} Hynes R.~I., Steeghs D., Casares J., Charles P.~A., O'Brien, K., 
           2003, ApJ, 583, L95 

\bibitem[Joinet et al.(2008)]{2008ApJ...679..655J} Joinet A., Kalemci E., Senziani F., 2008, ApJ, 679, 655

\bibitem[Kalemci et al.(2006)]{2006ApJ...639..340K} 
Kalemci E., Tomsick J.~A., Rothschild R.~E., Pottschmidt K., Corbel S., Kaaret P., 2006, ApJ, 639, 340

\bibitem[Kanbach et al.(2001)]{2001Natur.414..180K} Kanbach G., Straubmeier C., Spruit H.~C., Belloni, T., 2001, Nature, 414, 180 

\bibitem[Laurent \& Titarchuk (2001)]{2001ApJ...562L..67L} Laurent P., Titarchuk L., 2001, ApJ, 562, L67

\bibitem[Maccarone \& Coppi(2003)]{2003MNRAS.338..189M} Maccarone T.~J., Coppi P.~S., 2003, MNRAS, 338, 189

\bibitem[Maccarone(2003)]{2003A&A...409..697M} Maccarone T.~J., 2003, A\&A, 409, 697

\bibitem[Maccarone \& Schnittman(2005)]{2005MNRAS.357...12M} Maccarone T.~J., Schnittman J.~D., 2005, MNRAS, 357, 12

\bibitem[Malzac et al.(2004)]{2004MNRAS.351..253M} Malzac J., Merloni A., Fabian, A.~C., 2004, MNRAS, 351, 253 
 
 \bibitem[Markoff (2010)]{2010LNP...794..143M} Markoff S., 2010, in Belloni T., ed, 
       The jet paradigm: from  microquasars to quasars, Lecture Notes in Physics, Springer Verlag, Berlin, 794, 143

 \bibitem[Markoff et al.(2001)]{2001A&A...372L..25M} Markoff S., Falcke, H., Fender, R., 2001, A\&A, 372, L25

\bibitem[Markoff et al.(2002)]{2002APS..APRN17105M} Markoff S., Falcke  H.,  Fender, R., 2002, 
              APS Meeting Abstracts, 17105

\bibitem[McClintock \& Remillard(1986)]{1986ApJ...308..110M} McClintock J.~E.,  Remillard, R.~A., 1986, ApJ, 308, 110 

\bibitem[McClintock et al.(2011)]{2011CQGra..28k4009M} McClintock J.~E., et al., 2011, Classical and Quantum Gravity, 28, 114009 

\bibitem[Migliari et al.(2007)]{2007ApJ...670..610M} Migliari S., et al., 2007, ApJ, 670, 610

\bibitem[Miyamoto et al.(1992)]{1992ApJ...391L..21M} Miyamoto S., Kitamoto S., Iga S., Negoro H., Terada K., 1992, ApJ, 391, L21

 \bibitem[Miyamoto et al.(1993)]{1993ApJ...403L..39M} Miyamoto S., Iga S., Kitamoto S., Kamado Y., 1993, ApJ, 403, L39
 
 \bibitem[Morgan et al.(1997)]{1997ApJ...482..993M} Morgan E.~H., Remillard R.~A., Greiner J., 1997, ApJ, 482, 993 

  \bibitem[Motta et al.(2009)]{2009MNRAS.400.1603M} Motta S., Belloni T., Homan, J., 2009, MNRAS, 400, 1603
  
  \bibitem[Motta et al.(2010)]{2010MNRAS.408.1796M} Motta S., Mu{\~n}oz-Darias T., Belloni T., 2010, MNRAS, 408, 1796

\bibitem[Motta et al.(2011)]{2011arXiv1108.0540M} Motta S., Mu{\~n}oz-Darias T., Casella P., Belloni T., Homan J., 2011, MNRAS, in press (arXiv:1108.0540)

\bibitem[Mu{\~n}oz-Darias et al.(2007)]{2007MNRAS.379.1637M} Mu{\~n}oz-Darias T., Mart{\'{\i}}nez-Pais I.~G., 
 Casares J., Dhillon V.~S., Marsh T.~R., Cornelisse R., Steeghs D., Charles, P.~A., 2007, MNRAS, 379, 1637 

\bibitem[Mu{\~n}oz-Darias et al.(2008)]{2008MNRAS.385.2205M} Mu{\~n}oz-Darias T., Casares J., 
          Mart{\'{\i}}nez-Pais I.~G., 2008, MNRAS, 385, 2205 

  \bibitem[Mu{\~n}oz-Darias et al.(2010)]{2010MNRAS.404L..94M} Mu{\~n}oz-Darias T., Motta S., Pawar D., 
          Belloni T.~M., Campana S.,  Bhattacharya, D., 2010, MNRAS, 404, L94

\bibitem[Mu{\~n}oz-Darias et al.(2011a)]{2011MNRAS.410..679M} Mu{\~n}oz-Darias T., Motta S., Belloni, T.~M., 
         2011a, MNRAS, 410, 679

\bibitem[Mu{\~n}oz-Darias et al.(2011b)]{2011MNRAS.415..292M} Mu{\~n}oz-Darias T., Motta S., Stiele H.,  
           Belloni, T.~M.\ 2011b, MNRAS, 415, 292

\bibitem[Nespoli et al.(2003)]{2003A&A...412..235N} Nespoli E., Belloni T., Homan J., Miller J.~M., Lewin W.~H.~G., M{\'e}ndez M., van der Klis M., 2003, A\&A, 412, 235

\bibitem[Nowak \& Vaughan(1996)]{1996MNRAS.280..227N} Nowak M.~A., Vaughan B.~A., 1996, MNRAS, 280, 227

\bibitem[Orosz et al.(2007)]{2007Natur.449..872O} Orosz J.~A., et al., 2007, Nature, 449, 872 

\bibitem[{\"O}zel et al.(2010)]{2010ApJ...725.1918O} {\"O}zel F., Psaltis D., Narayan R., McClintock, J.~E., 
            2010, ApJ, 725, 1918 

\bibitem[Rao et al.(2010)]{2010ApJ...714.1065R}  Rao F., Belloni T., Stella L., Zhang S.~N., Li T., 2010, ApJ, 714, 1065

 \bibitem[Reig, Belloni, \& van der Klis(2003)]{2003A&A...412..229R} Reig P., Belloni T., van der Klis M., 2003, A\&A, 412, 229

 \bibitem[Remillard \& McClintock(2006)]{2006ARA&A..44...49R} Remillard R.~A., McClintock J.~E., 2006,
ARA\&A, 44, 49
  
  \bibitem[Remillard et al.(2002)]{2002ApJ...564..962R} Remillard R.~A., Sobczak G.~J., Muno M.~P., McClintock J.~E., 2002, ApJ, 564, 962
  
  \bibitem[Romani(1998)]{1998A&A...333..583R} Romani R.~W., 1998, A\&A, 333, 583 
  
  \bibitem[Russell et al.(2006)]{2006MNRAS.371.1334R} Russell D.~M., Fender R.~P., Hynes R.~I., Brocksopp C., Homan J., Jonker P.~G., Buxton M.~M., 2006, MNRAS, 371, 1334
  
  \bibitem[Russell et al.(2010)]{2010MNRAS.405.1759R} Russell D.~M., Maitra D., Dunn R.~J.~H., Markoff, S., 2010, MNRAS, 405, 1759 
  
  \bibitem[Russell et al.(2011)]{2011arXiv1106.0723R} Russell D.~M., Miller-Jones J.~C.~A., Maccarone T.~J., Yang Y.~J., Fender R.~P., Lewis F., 2011, MNRAS, in press (arXiv:1106.0723)
  
  \bibitem[Soleri, Belloni, \& Casella(2008)]{2008MNRAS.383.1089S} Soleri P., Belloni T., Casella P., 2008, MNRAS, 383, 1089
  
  \bibitem[Stiele et al.\ (2011)]{2011arXiv1108.2198S}  Stiele H., Motta S., Mu{\~n}oz-Darias T., Belloni T.~M., 2011, MNRAS, in press (arXiv:1108.2198)
  
  \bibitem[Takizawa et al.(1997)]{1997ApJ...489..272T} Takizawa M., et al., 1997, ApJ, 489, 272

 \bibitem[Tanaka \& Lewin(1995)]{1995xrbi.nasa..126T} 
Tanaka Y., Lewin W.~H.~G., 1995, in  Eds.  Lewin W.H.G., van Paradijs J., van den Heuvel E.P.J., eds, X-ray
binaries, Cambridge Univ. Press, p. 126

\bibitem[Uttley et al.(2011)]{2011MNRAS.414L..60U} Uttley P., Wilkinson
T., Cassatella P., Wilms J., Pottschmidt K., Hanke M., Bock M., 2011, MNRAS, 414, L60

\bibitem[van der Klis(2006)]{2006csxs.book...39V} van der Klis M., 2006, in Lewin W., van der Klis M., eds.,
         Compact stellar X-ray sources, Cambridge University Press, Cambridge, p. 39

\bibitem[Veledina et al.(2011)]{2011ApJ...737L..17V} Veledina A., Poutanen J., Vurm, I., 2011, ApJ, 737, L17 

 \bibitem[Wijnands \& van der Klis(1999)]{1999ApJ...514..939W} Wijnands R., van der Klis, M., 1999, ApJ, 514, 939
 
 \bibitem[Wijnands et al.(1999)]{1999ApJ...526L..33W} Wijnands R., Homan J., van der Klis M., 1999, ApJ, 526, L33

\bibitem[Wilkinson \& Uttley(2009)]{2009MNRAS.397..666W} Wilkinson T., Uttley P., 2009, MNRAS, 397, 666

\bibitem[Yu \& Yan(2009)]{2009ApJ...701.1940Y} Yu W., Yan Z., 2009, ApJ, 701, 1940

\bibitem[Zdziarski et al.(2002)]{2002ApJ...578..357Z} Zdziarski A.~A., Poutanen J., Paciesas W.~S.,  Wen, L., 2002, ApJ, 578, 357

\end{thebibliography}
\end{document}